\documentclass[a4paper]{jpconf}  
\usepackage[dvips]{graphicx}
\usepackage{dcolumn}   
\usepackage{bm}        
\usepackage{amssymb}   
\usepackage[utf8]{inputenc}
\usepackage{mathrsfs}
\usepackage{amsmath}
\usepackage[english]{babel}
\usepackage{hyperref}
\usepackage{textcomp}
\usepackage{subfigure}
\usepackage{color}
\usepackage{cite}
\def\eqref#1{(\ref{#1})}

\def\bfl{\begin{flushleft}}
\def\efl{\end{flushleft}}
\def\bfr{\begin{flushright}}
\def\efr{\end{flushright}}
\def\bc{\begin{center}}
\def\ec{\end{center}}
\def\be{\begin{equation}}
\def\ee{\end{equation}}
\def\bse{\begin{subequations}}
\def\ese{\end{subequations}}
\def\ba{\begin{eqnarray}}
\def\ea{\end{eqnarray}}
\def\baa#1{\begin{array}{#1}}
\def\eaa{\end{array}}
\def\bw{\begin{widetext}}
\def\ew{\end{widetext}}

\def\bit{\begin{itemize}}
\def\eit{\end{itemize}}
\def\bco{}
\def\bcs{\begin{cases}}
\def\ecs{\end{cases}}

\begin{document}

\title{Kinks in the relativistic model with logarithmic nonlinearity}

\author{E.~Belendryasova$^{1}$, V.~A.~Gani$^{1,2}$ and K.~G.~Zloshchastiev$^3$}

\address{$^1$National Research Nuclear University MEPhI (Moscow Engineering Physics Institute), 115409 Moscow, Russia}

\address{$^2$National Research Center Kurchatov Institute, Institute for Theoretical and Experimental Physics, 117218 Moscow, Russia}

\address{$^3$Institute of Systems Science, Durban University of Technology, P.O. Box 1334, Durban 4000, South Africa}

\ead{vagani@mephi.ru}

\begin{abstract}
We study the properties of 
a relativistic model with logarithmic nonlinearity. 
We show that such model allows two types of solutions: topologically trivial (gaussons) and topologically non-trivial (kinks), depending on a sign of the nonlinear coupling.  
We focus primarily on the kinks' case and study their scattering properties.
For the kink-antikink scattering, we have found a critical value of the initial velocity,
which separates two different scenarios of scattering. 
For the initial velocities 
below this critical value,
the kinks form a bound state, which then decays slowly. 
If 
the initial velocities are
above the critical value,
the kinks collide, bounce and eventually escape to infinities. During this process, the higher initial velocity is, the greater is the elasticity of the collision. We also study excitation spectrum of the kink solution.
\end{abstract}

\section{Introduction}

Kinks are a very important class of topologically non-trivial solitonic solutions of field-theoretic models in $(1+1)$ dimensions. Studies of the kink-(anti)kink interactions and related resonance phenomena is a fast growing area of research, where many important results have been obtained recently. In particular, the kink-antikink scattering has been studied within models with both polynomial \cite{Gani.YaF.2001.SuSy.eng,Gani.YaF.2001.SuSy.rus,Dorey.PRL.2011,Gani.PRD.2014,Gani.JHEP.2015,Gani.IOP.2016,Belendryasova.CNSNS.2019,Belendryasova.IOP.2017.phi8} and non-polynomial potentials \cite{Gani.PRE.1999,Bazeia.EPJC.2018.sinh,Bazeia.JPCS.2017.sinh,Gani.EPJC.2018.dsg,Belendryasova.2018}. The role of collective excitations of the kink-antikink system in appearance of the escape windows has been clarified \cite{Gani.PRE.1999,Dorey.PRL.2011,Belendryasova.CNSNS.2019,Belendryasova.IOP.2017.phi8}, the impact of quasi-normal modes on the resonance phenomena has been shown \cite{Dorey.PLB.2018}. A series of papers is devoted to multi-kink collisions in various models \cite{Moradi.JHEP.2017,Moradi.EPJB.2017,Moradi.CNSNS.2017,Gani.arXiv.2019.DSG.multi}. Another important issue which is being actively studied is interaction of kinks with polynomial asymptotic behavior. It has been shown that polynomial tails result in long range interactions of kinks, see, e.g., \cite{Radomskiy.JPCS.2017,Belendryasova.CNSNS.2019,Bazeia.JPhCom.2018,Christov.2018.long-range,Christov.2018.long-tails,Manton.2018.forceK,Manton.2018.forceKA}.

This our paper deals with scattering of kink and antikink of the relativistic model with logarithmic nonlinearity. The latter is being actively studied starting from the classical works by Rosen and Bialynicki-Birula and Mycielski \cite{ros68,ros69,bb75,bb76,bbm79}. Currently they find numerous applications in different areas of physics, including the  particle physics and classical and quantum gravity \cite{ros68,ros69,em98,hkt10,z10gc,z11ap,z11pla,dz11,gul14,gul15,dmz15,szm16}. Modern statistical mechanical justifications of logarithmic models and their connections to fundamental quantum-mechanical notions can be found in \cite{z18zna},
and their applicability to other branches of physics is discussed in \cite{zrz17,z18epl}.



\section{The model}

Consider a real scalar field $\phi$ in $(1+1)$-dimensional space-time with its dynamics defined by the Lagrangian \cite{ros68,ros69}
\begin{equation}\label{eq:lagrangian}
\mathcal{L} = \frac{1}{2}\left(\frac{\partial \phi}{\partial t}\right)^2-\frac{1}{2}\left(\frac{\partial \phi}{\partial x}\right)^2-V(\phi),
\end{equation}
with the potential
\begin{equation}\label{eq:pot_gen}
V (\phi) =-b\, \phi^2 \left[\ln{({\cal A} \phi^2) - 1}\right]+ V_0,
\end{equation}
where the parameter $b$ defines the strength of a nonlinear self-interaction, $\cal{A}$ is a constant parameter, and $V_0 = V (0)$ is the value of the  potential at $\phi = 0$ \cite{z11ap}. From the Lagrangian \eqref{eq:lagrangian} we can obtain the equation of motion
\begin{equation}\label{eq:eqmo}
\frac{\partial^2\phi}{\partial t^2}-\frac{\partial^2\phi}{\partial x^2}+\frac{dV}{d\phi}=0.
\end{equation}
In static case $\phi=\phi(x)$:
\begin{equation}\label{eq:eqmo_stat}
\frac{d^2\phi}{dx^2}=\frac{dV}{d\phi}\quad\Leftrightarrow\quad \frac{d\phi}{dx}=\pm\sqrt{2V}.
\end{equation}

\paragraph{ (a) Gausson Model.} Assume that $b>0$ and $V_0=0$, therefore the potential~\eqref{eq:pot_gen} can be rewritten:
\begin{equation}\label{eq:potgau}
V^{(g)} (\phi) =- \frac{\phi^2 }{\lambda^2}\left[\ln{(\phi^2/\vartheta^2) - 1}\right],
\end{equation}
where $\lambda =1/{\sqrt{b}}$ and ${\cal A} =1/\vartheta^2$ are real parameters.
In this case the field potential has a shape of an upside-down  Mexican hat placed inside a well with walls representing
the cut-off value of $|\phi|$. Then from equation~\eqref{eq:eqmo_stat} we can obtain static solutions known as gaussons \cite{ros68,bbm79}:
\begin{equation}\label{eq:gau}
\phi^{(g)}_{\pm} =\phi_0\exp\left[- \frac{x (x \pm \mu)}{\lambda^2}\right],
\end{equation}
where 
$\mu = \sqrt{2}  \lambda \sqrt{1 - \ln{(\phi_0^2/\vartheta^2)}}$. Their energy density and total mass-energy are 
\begin{equation}\label{eq:Hgau}
\mathcal{H}^{(g)}_\pm =\left[\frac{2}{\lambda^2}\left(x \pm \frac{\mu}{2}\right)\phi^{(g)}_\pm\right]^2=\frac{4 \phi_0^2}{\lambda^4}\left(x \pm \frac{\mu}{2}\right)^2\exp{\left[- \frac{2 x (x \pm \mu)}{\lambda^2}\right]},
\end{equation}
\begin{equation}\label{eq:Egau}
E^{(g)}_\pm \equiv E \left[\phi^{(g)}_\pm \right] =\sqrt{\frac{\pi}{2}}\frac{\text{e} \vartheta^2}{\lambda}.
\end{equation}

\paragraph{ (b) Kink Model.}

Consider a case with $b < 0$ and $V_0 = - b$,
therefore the potential~\eqref{eq:pot_gen} can be rewritten as:
\begin{equation}
V^{(k)} (\phi) =\frac{\phi^2}{\ell^2}\left[\ln{(\phi^2/\vartheta^2) - 1}\right]+\frac{\vartheta^2}{\ell^2},
\end{equation}
where $\ell = 1/\sqrt{-b}$ is a real parameter.
In this case, the potential has a conventional Mexican-hat shape thus hinting at
the presence of nontrivial topological structure \cite{z11ap}.
Unlike the gausson case, in this case static solutions are obtained so far only numerically.


\section{Kinks scattering}

We have studied the kink-antikink scattering, i.e.\ we have taken the kink and antikink, initially (at the moment $t=0$) located at the points $x=-\xi$ and $x=+\xi$, respectively, and moving towards each other with initial velocities $v_\mathrm{in}$ in the laboratory frame of reference:
\begin{equation}
\phi(x,t) = \phi_{\rm kink}\left(\frac{x+\xi-v_\mathrm{in}t}{\sqrt{1-v_\mathrm{in}^2}}\right)+\phi_{\rm antikink}\left(\frac{x-\xi+v_\mathrm{in}t}{\sqrt{1-v_\mathrm{in}^2}}\right)-1
\end{equation}
(we used parameters of potential $\vartheta=1, \ell=\sqrt{2}$.) We solved the equation~\eqref{eq:eqmo} numerically via explicit difference scheme
\begin{equation}
\frac{\partial^2\phi}{\partial t^2} = \frac{11\phi_{n}^{j+1} -20\phi_{n}^{j}+6\phi_{n}^{j-1}+4\phi_{n}^{j-2}-\phi_{n}^{j-3}}{12\delta t^2},
\end{equation}
\begin{equation}
\frac{\partial^2\phi}{\partial x^2} = \frac{-\phi_{n-2,j} +16\phi_{n-1}^{j}-30\phi_{n}^{j}+16\phi_{n+1}^{j}-\phi_{n+2}^{j}}{12\delta x^2}
\end{equation}
with time step $\delta t=0.001$ and coordinate step $\delta x=0.005$.

The scattering process crucially depends on the initial velocity of the colliding kinks. At low energies we observed the kinks' capture and formation of their bound state which then decays into radiation. At the same time, at high energies the kinks bounce off each other and goes to spatial infinities. In our numerical experiments we have found a critical value $v_\mathrm{cr}\approx 0.7908$ of the initial velocity which separates two regimes of scattering: at $v_\mathrm{in} < v_\mathrm{cr}$ kinks form a bound state, while at $v_\mathrm{in} > v_\mathrm{cr}$ the kinks collide once and escape to infinities with final velocities $v_\mathrm{f} < v_\mathrm{in}$.

Note that, unlike the $\phi^4$ and many other models, in the kink-antikink collisions at $v_\mathrm{in} < v_\mathrm{cr}$ we did not observe formation of a bion -- a localized long-lived bound state of kink and antikink. In our numerical simulations the kink and antikink captured each other and formed a bion-like state which then immediately decayed into radiation.


\section{Discussion}

We have studied the scattering of kinks of the relativistic model with logarithmic nonlinearity. We have found a critical value of the initial velocity of the colliding kinks $v_\mathrm{cr}\approx 0.7908$, which separates different regimes of the scattering. At the initial velocities below $v_\mathrm{cr}$ the kink and antikink become trapped thus forming a bound state. This bound state then decays into radiation in the form of waves of small amplitude. At the initial velocities above $v_\mathrm{cr}$, the kink and antikink collide once and escape to infinities with some final velocities $v_\mathrm{f}<v_\mathrm{in}$, i.e., a kind of inelastic scattering takes place. A certain part of the initial kinetic energy of kinks is emitted in a form of waves of a small amplitude.

It is worth mentioning that we did not observe any resonance phenomena (so-called 'escape windows') so far. It can be a consequence of absence of vibrational mode(s) in the kink's excitation spectrum. In order to confirm this conjecture, we are planning study the linear stability of the kink in a spirit of refs.~\cite{Gani.PRD.2014,Gani.JHEP.2015,Belendryasova.CNSNS.2019}. 


\section*{Acknowledgments}
The work of MEPhI group was supported by the MEPhI Academic Excellence Project (Contract No.\ 02.a03.21.0005, 27.08.2013). E.B.\ and V.A.G.\ also acknowledge the support of the Russian Foundation for Basic Research under Grant No.\ 19-02-00971. The work of K.Z.\ is supported by the National Research Foundation of South Africa under Grants Nos.\ 95965 and 98892. Numerical simulations were performed using resources of NRNU MEPhI high-performance computing center.

\section*{References}

\bibliography{my_biblio,biblio2}

\end{document}